\documentclass{cernyrep}
\usepackage[T1]{fontenc}
\usepackage[bookmarks, colorlinks=true, linktoc=page, pdftex, linkcolor=black, citecolor=black, urlcolor=blue]{hyperref}

\usepackage{fancyhdr}
\fancyhfoffset{4 mm}
\fancypagestyle{ARTTITLE}{%
\fancyhf{} 
\lhead{\raggedright {\it Mechanical \& Materials Engineering for Particle Accelerators and Detectors}\raggedright \\ CERN Accelerator School Proceedings ---  Sint-Michielsgestel,  Netherlands, 2024\hfill}
\lfoot{\hspace{3mm} Available online at \url{https://cas.web.cern.ch/previous-schools}}
\rfoot{\thepage\hspace*{3mm}}
 
}

\usepackage{varwidth}
\usepackage{xcolor}
\begin{document}

\title{RF Applications}
\author{Thomas Geoffrey Lucas}
\institute{Paul Scherrer Institut, Villigen, Switzerland}

\begin{abstract}
Radiofrequency (RF) systems play a critical role in particle accelerators by enabling the acceleration, manipulation, and diagnosis of charged particle beams. At the heart of many of these systems lies the RF cavity, whose effective design requires close collaboration among RF designers, beam physicists, and mechanical engineers. This chapter presents the fundamental principles of RF systems, with particular emphasis on RF cavities, and underscores the interdisciplinary effort involved in their development. The SwissFEL X-ray free-electron laser at the Paul Scherrer Institut serves as a key example to illustrate these concepts.
\end{abstract}
\keywords{Radiofrequency; Accelerating Structures, RF Design, Mechanical Engineering.}
\maketitle
\thispagestyle{ARTTITLE}

\section{Introduction}
Radiofrequency (RF) is fundamental to particle accelerators. By storing RF in cavities, accelerators can impart kinetic energy ranging from the electron-Volt (eV) to Teraelectron-Volt ($10^{12}$~eV) scale~\cite{LHC}. While its primary role is beam acceleration, RF is also essential for beam diagnostics, post beam-generation acceleration, and beam manipulation.

This chapter explores key RF applications and highlights the role of mechanical engineering in developing RF systems, in particular for the design and realisation of the electromagnetic cavity, which from now on we'll refer to as the `RF cavity'. We begin with background theory and commonly used terminology, then examine the functionality of RF cavities and transmission lines, which are core elements in many RF systems. The mechanical engineer's contributions are discussed throughout, emphasising interdisciplinary collaboration. Finally, we illustrate these concepts using the SwissFEL accelerating structure as a detailed case study. The reader should note that here we will use `cavity' to describe a single electromagnetic resonator or multiple electromagnetic resonators that resonate together, the `cell' to describe a repeated geometric structure whose shape forms cavities and the term `accelerating structure' will refer to a concatenated group of cavities.

\section{Fundamental Concepts in RF Accelerator Physics}

\begin{figure}
    \centering
    \includegraphics[width=0.75\linewidth]{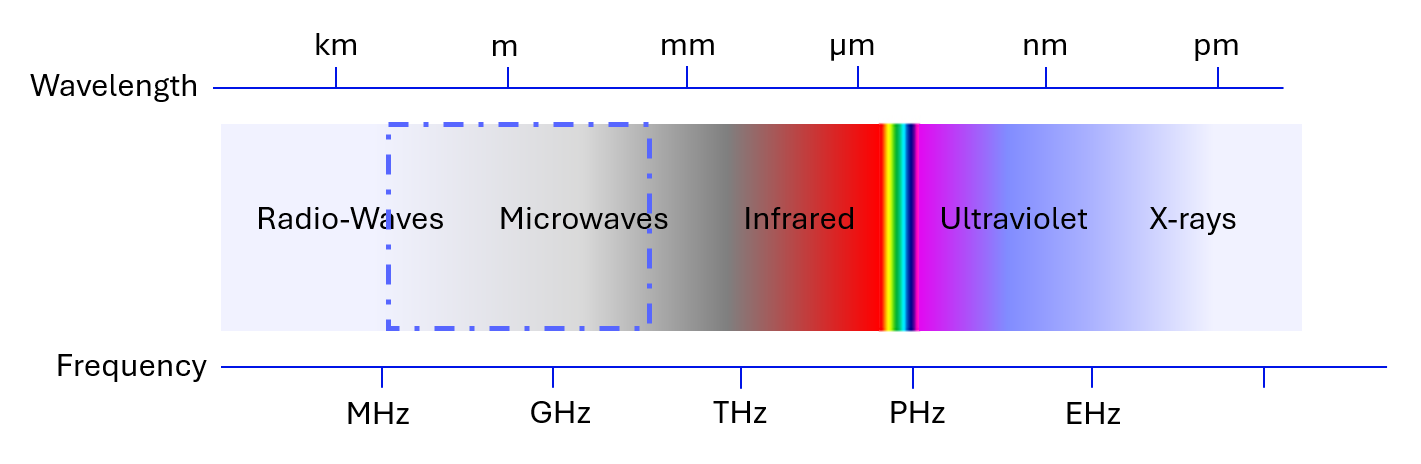}
    \caption{The electromagnetic spectrum and region of whose frequencies are commonly used in RF cavities. Frequencies below 3 MHz can be generated by other RF systems in accelerators, such as beam diagnostics.}
    \label{fig:fig1}
\end{figure}

\begin{figure}
    \centering
    \includegraphics[width=0.75\linewidth]{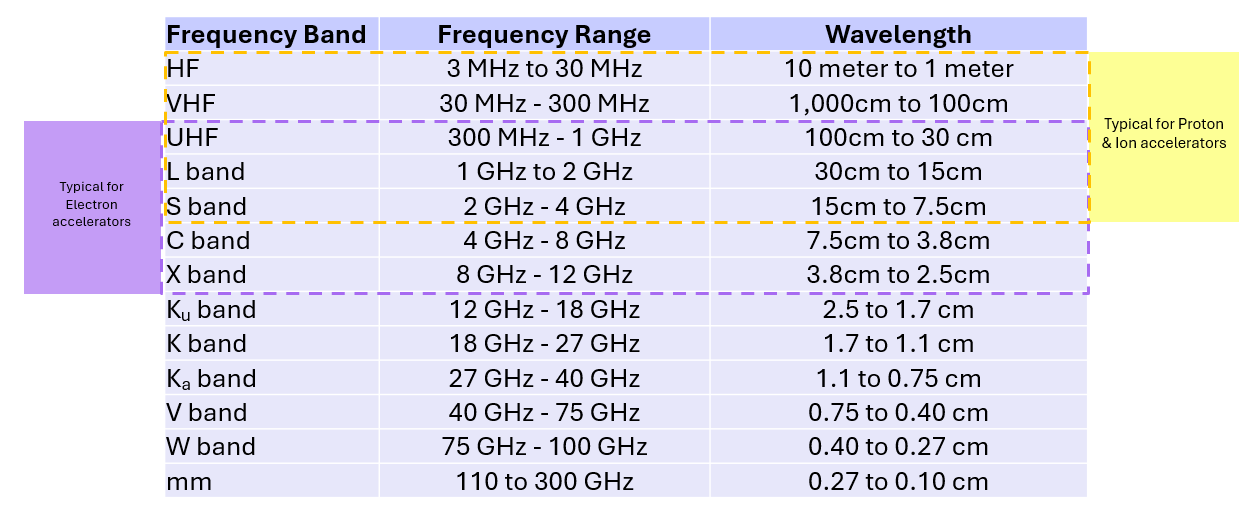}
    \caption{Frequency bands of the Radiofrequency and Microwave region of the electromagnetic spectrum.}
    \label{fig:fig1a}
\end{figure}

RF systems are grounded in electromagnetism, specifically in the few Hz to $\sim$300 GHz range of the~spectrum (see Fig.~\ref{fig:fig1}). This includes the radiowave and microwave bands, which are the basis of RF technology and the former of which lends its name to RF systems. Within this spectrum, it is common to discuss the technology within certain frequency bands (see Fig.~\ref{fig:fig1a}). The use of each frequency band in RF technology depends on its application. For accelerators, it is common that electron accelerators use higher frequencies than proton accelerators. But the choice in frequency also depends on, among many other factors, whether there is the need to have continuous or pulsed operation and the machine size. Before exploring RF systems in detail, we briefly review key electromagnetic principles and understand how they relate to the acceleration of charged particles.

\subsection{Interactions of Electromagnetic Fields and Charged Particles}

Electromagnetism consists of coupled electric and magnetic fields, which can interact with charged particles, a principle central to accelerator physics. The force on a charged particle in an electromagnetic field is given by the Lorentz force:
\begin{equation}
    \vec{F} = q(\vec{E} + \vec{v} \times \vec{B}) .
    \label{Eqn:Eqn1}
\end{equation}
Here, \( q \) is the particle charge, \( \vec{E} \) the electric field, \( \vec{B} \) the magnetic field, and \( \vec{v} \) the particle velocity. Electric fields exert force in the direction of the field, enabling acceleration. Magnetic fields, by contrast, act perpendicular to the particle’s velocity and field direction, making them suitable for beam steering and focusing. 

However, Eq.~(\ref{Eqn:Eqn1}) shows that it would also be possible to use electric fields to steer. To illustrate why, in most situations, we use magnetic fields for steering and focusing rather than electric fields we take the example of an `ultrarelativistic' particle (defined as a particle whose velocity can be assumed to be approximately the speed of light \( v \approx c \)). Such a charged particle experiences the same transverse force from either a 1~T magnetic field or a 299.8~MV/m electric field when each is applied perpendicular to its velocity. The magnetic option is far more practical, as electric fields at this scale exceed the~limits of current continuous-power technologies while such magnetic field strengths are achievable with current technology.\\

\begin{figure}
    \centering
    \includegraphics[width=0.75\linewidth]{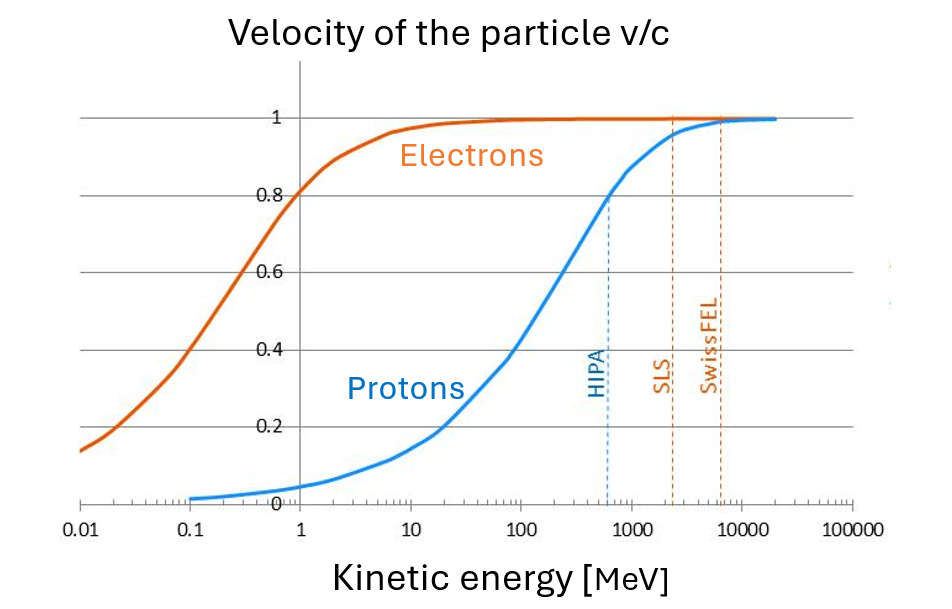}
    \caption{Velocity vs kinetic energy for electron and protons. The three large-scale facilities of PSI are illustrated with the peak operational energy.}
    \label{fig:fig2}
\end{figure}

\subsection{Key Concepts in Accelerator Physics}
Understanding special relativity and accelerator physics is essential for RF applications in accelerators, though in most cases, mechanical engineers only need a basic familiarity with key concepts and terminology to effectively communicate with RF and beam dynamics specialists.

A commonly used unit in accelerator physics is the electron-volt (eV), which describes the kinetic energy of individual charged particles. Unlike the Joule, which is too large for this scale, the eV is convenient for describing this energy scale as 1~eV is the energy gained by an electron when accelerated through a potential of 1~Volt.

A key concept in RF systems of accelerators is the relationship between particle velocity (\(v\)) and kinetic energy (\(K\)). As shown in Fig.~\ref{fig:fig2}, velocity does not continue increasing classically (\(v = \sqrt{2K/m}\)). Instead, due to relativistic effects, particles approach a maximum velocity of the speed of light (\(c\)), asymptotically. This dynamic occurs at different energy scales for different particles due to their difference in mass. Electrons, being roughly 2000 times lighter than protons, reach near-light speeds within a few meters of acceleration after their generation, given a typical particle accelerator. However, protons require much longer distances to approach $c$, often hundreds of meters to kilometers of acceleration. The particle velocity has important implications for RF cavity design: the length of an RF cavity (in the direction of travel for the~beam) should relate to the particle velocity and the frequency of the RF to achieve efficient acceleration.

\section{RF Systems}

With a few key physics principles described, we now turn to RF systems themselves. This section focuses on how RF is used, particularly in accelerating structures and power transmission, and highlights the role of mechanical engineers in designing these systems.

\subsection{Applications in Accelerators}

RF systems are essential to many functions in particle accelerators. Most notably, they are used for the: 
\begin{itemize}
\item \textbf{Acceleration} of the particle bunch(es);
\item \textbf{Manipulation} of the particle bunch(es): To compress, stretch, rotate, or linearise;
\item \textbf{Diagnosis} of the particle bunch's properties.
\end{itemize}
For brevity, we will focus on the primary application of acceleration, and the RF cavities used, where careful mechanical engineering is key to their successful operation.\\

\subsection{RF Power Transmission: Coaxial Lines and Waveguides}

The first step in any RF system, after the generation of the RF power, is its transmission. Two common technologies for RF transmission are:

\textbf{Coaxial Cables}: These consist of an inner conductor and an outer conductor separated by a dielectric. These cables are widely used due to their efficacy across a broad range of frequencies. However, their use at high RF power is limited, as the dielectric makes it difficult to cool the inner conductor.

\textbf{Waveguides}: These are hollow, pipe-like structures that transmit electromagnetic power. Waveguides are more widely used at high power due to their low-loss nature and simplicity to cool. However, a given waveguide geometry only operates over a limited range of frequencies whose geometry grows inverse to the RF frequency. Consequently, waveguides become large and unwieldy when used at low frequencies, making them impractical in some cases. To give an example, a waveguide is used in a 500~MHz high-power RF system; however, its width is 0.45 m.

In both coaxial cables and waveguides, electromagnetic energy is transmitted in specific patterns known as \textbf{modes}. In waveguides, common modes include \textit{transverse electric (TE)} and \textit{transverse magnetic (TM)}. In coaxial cables, the primary mode is the \textit{transverse electromagnetic (TEM)}.

\subsection{RF Cavities and Acceleration Modes}

The core component of an accelerator is the \textbf{RF cavity}, which stores electromagnetic energy and generates strong electric and magnetic fields used for accelerating or manipulating charged particle beams. The resonant frequency of a cavity depends on its geometry and supported mode, similar to a waveguide.

A commonly used mode is the \textbf{TM\textsubscript{010}} mode also known as the acceleration mode\footnote{Officially it is the TM\textsubscript{010}-like mode in an accelerating cavity as it is not a pure TM\textsubscript{010}}. The frequency of this mode in a perfect, closed cylindrical cavity is given by~\cite{Gerigk2011}:
\begin{equation}
    f_0 = \frac{2.40483\, c}{2 \pi a},
\end{equation}
where \( a \) is the radius of the cavity and \( c \) is the speed of light. As this equation shows, smaller RF cavity radii are needed for higher frequencies. Therefore, we see that the cavity's internal dimensions relate to the RF frequency (through its radius) and the charged particle's velocity (through its length).

\subsection{Common Nomenclature in RF and Figures-of-Merit}

Previously, we discussed the relationship between electric and magnetic fields, and their impact on the~beam. However, we have not yet described how these electromagnetic fields are generated or the~energy associated with them. A key concept in RF systems concerns the power required to generate and sustain the fields in cavities. The relationship between electromagnetic fields and power is characterised by several fundamental parameters, commonly used by RF designers. These concepts are introduced below.

Power loss in an RF system arises from the finite conductivity of the cavity walls. The resistivity ($\rho$) of the wall material defines the \emph{skin depth}:
\[
\delta = \sqrt{\frac{\rho}{\pi f_0 \mu_r \mu_0}},
\]
where $f_0$ is the RF frequency, $\mu_r$ is the relative permeability, and $\mu_0$ is the permeability of free space. This skin depth describes how far the electromagnetic fields permeate into the surface of the cavity. The~\emph{surface resistance} is then defined as
\[
R_s = \frac{\rho}{\delta}.
\]
With $R_s$ and the transverse surface magnetic field $H_t$, the power loss can be expressed as
\[
P_{\text{loss}} = \iint_{\text{wall}} R_s |H_t|^2 \, dA,
\]
where the integral is applied over the entire cavity wall. This calculation is important for mechanical engineers as it tells one where the sources of heating will come from. This will be illustrated with an~example in Section~\ref{Sec:Mech_design}.

Another important parameter derived from the power loss is the (unloaded) quality factor $Q$, which measures the ratio of stored energy in the cavity to power dissipated per cycle:
\[
Q = \frac{2 \pi f_0 W}{P_{\text{loss}}},
\]
where the stored energy ($W$) is
\[
W = \iiint \left( \frac{\epsilon}{2} |\vec{E}|^2 + \frac{\mu}{2} |\vec{H}|^2 \right) dV,
\]
and $\epsilon = \epsilon_r \epsilon_0$ is the permittivity of the cavity volume and $\mu = \mu_r \mu_0$ is the permeability of the cavity volume.
For copper cavities, appropriate $Q$ values generally lie in thousands or greater.

Another useful quantity is the \emph{shunt impedance}, which quantifies the voltage generated for a given power loss:
\[
R = \frac{|V_{\text{acc}}|^2}{2 P_{\text{loss}}},
\]
where $V_{\text{acc}}$ is the accelerating voltage. To describe the cavity geometry independent of material properties, one can define the ratio
\[
\frac{R}{Q} = \frac{|V_{\text{acc}}|^2}{4 \pi f_0 W}.
\]
It is common that an RF designer will optimise these quantities during the early stages of the design process,  where they primarily consider the internal (RF) geometry. It is important that the mechanical engineer follows this design to ensure that the RF design is also mechanically feasible.

\section{Common Mechanical Engineering Considerations}
Below we shift towards the mechanical engineering aspects continuing to focus on RF cavities.

\subsection{Material Choice}
When selecting the material for RF cavities the most common choice for those working with normal conducting materials is copper. Copper has many qualities desired by RF designers, for example it has:
\begin{itemize}
    \item very good electrical and thermal conductivity;
    \item good machinability, with the ability to achieve roughness levels in the nm-regime when using UP machining techniques;
    \item the ability to be brazed;
    \item good availability and cost;
    \item a low secondary emission yield value, important for multipacting mitigation.
\end{itemize}
In particular, the copper used by PSI to create our RF cavities in the GHz-regime is oxygen-free, high conductivity (Cu-OFE) with impurities in accordance with ISO 431. In addition to this, the Cu-OFE is 3D forged in order to remove voids within the material. Such voids may only become visible during the~final machining process or even exposed during high power operation.

In certain circumstances, the material choice can vary from copper. Some examples of this are: Niobium for superconducting cavities~\cite{SCmaterials}, CuAg alloy for very high electric field applications~\cite{CuAg}, or aluminium for a more lossy cavity~\cite{Alcavity}.

\begin{figure}
    \centering
    \includegraphics[width=0.45\linewidth]{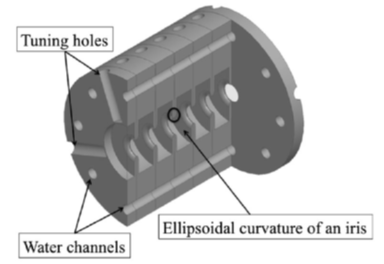}
    \includegraphics[width=0.45\linewidth]{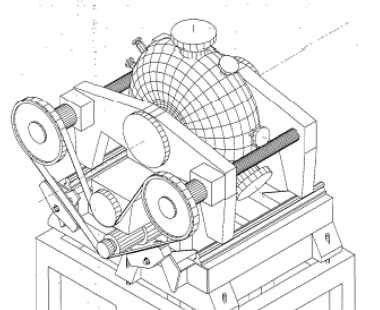}
    \caption{Tuning systems for a 5.712 GHz `high frequency' accelerating structure (left) and 500 MHz `low frequency' SLS Cavity (right). The high frequency device uses tuning pins placed inside holes to plastically deform the cavities individually while the low frequency SLS cavity uses an active tuning system to tune the single cavity through elastic deformations.}
    \label{fig:fig5}
\end{figure}

\subsection{Machining Tolerances}
A key question in the mechanical design of RF cavities is: \textit{what machining precision is required?} While we can not provide an exact number for all applications, as it depends on various parameters, a rule of thumb is: \textit{higher-frequency cavities require tighter mechanical tolerances.}
Other factors that influence the tolerance requirements include:
\begin{itemize}
    \item group velocity of structure, defined as the velocity of the power flow;
    \item quality factor of resonator;
    \item use of post-fabrication tuning.
\end{itemize}

\subsection{Joining Technologies}
After machining the individual components, these must be joint into the final device. Such joining aims to:
\begin{itemize}
    \item provide mechanical rigidity and maintain the vacuum volume geometry given by the RF designer;
    \item  separate the air, water and vacuum sections. Possibly being able to maintain ultrahigh vacuum (if necessary);
    \item allow heat transfer and surface currents to flow between the joint pieces (if required).
\end{itemize}
The technologies to join components are wide-spanning including, among other techniques: brazing, diffusion bonding, electron beam welding, clamping with gaskets, and TIG welding.
The choice of material depends on the application, as described above, which affects the suitability of the joining method. For example, high-gradient cavities may use soft or hard copper. Soft copper can be brazed or diffusion bonded, while hard copper should avoid excessive heating (to prevent softening); thus, clamping or welding is preferred.

\subsection{Post-fabrication Tuning of Cavities}
In instances where it is difficult to achieve the machining tolerance or required dimensions after joining, a post-fabrication tuning is required. Such tuning may also be important if one wants to actively (de)tune the cavity during operation.

It is important to consider how to tune such cavities. It is common to use the following methods, based on frequency:
\begin{itemize}
    \item Low Frequency RF Cavities: Elastic deformation and plungers;
    \item High Frequency RF Cavities: Plastic deformation and temperature changes.
\end{itemize}
Figure~\ref{fig:fig5} depicts the design of two RF devices at 5.712~GHz (left) and 500~MHz (right) and their associated tuning systems.
The reason for the use of temperature at higher frequencies and not at lower frequencies relates to the change in frequency per unit temperature. The change in cavity frequency for a given temperature change is:
\begin{equation}
    \Delta f = \alpha f \Delta T,
\end{equation}
where $\alpha$ is the thermal expansion coefficient, $\Delta T$ is the temperature difference from the nominal temperature and $f$ is the frequency of the cavity. For C-band cavities at 5.712~GHz, this equates to 97 kHz/$^\circ$C while at 500 MHz this is only 8.5 kHz/$^\circ$C.

\section{Example: Design and Realisation of an RF Accelerating Structure for SwissFEL}
Using the knowledge that has been described, this section aims to overview the process of designing an~accelerating structure for SwissFEL. We begin with the electromagnetic design then move on to the~mechanical design, realisation and validation of its performance.

\subsection{Electromagnetic Design}
\begin{figure}
    \centering
    \includegraphics[width=0.9\linewidth]{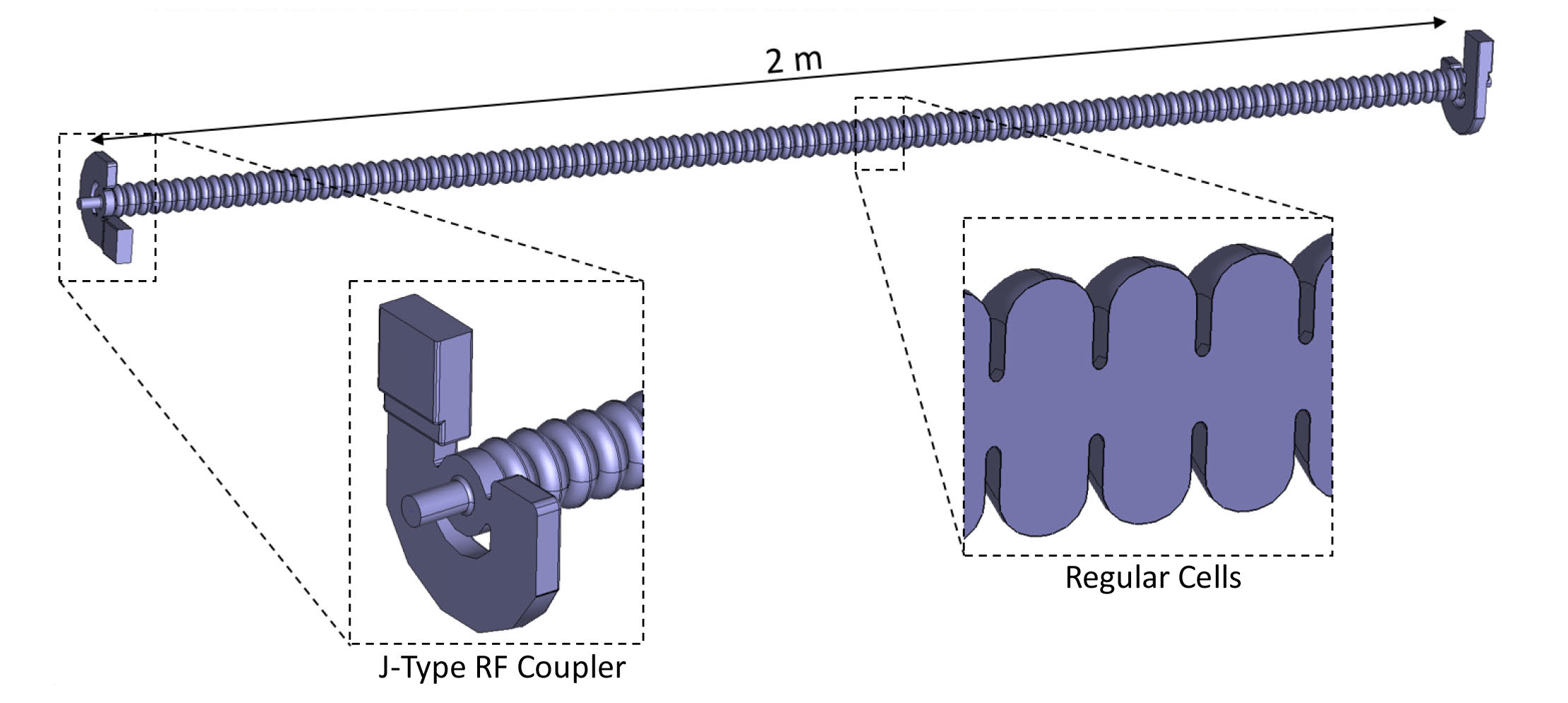}
    \caption{Electromagnetic design of the SwissFEL C-band accelerating structure~\cite{Lucas2025}}
    \label{fig:fig6}
\end{figure}

Beginning with the RF cavity design, followed by a concatenation of these into a full accelerating structure with input/output coupler(s), the aim of this step is to optimise the electromagnetic properties of the accelerating structure. This will include beam dynamics and the RF power usage considerations. Although not commonly being directly involved in the RF design, during this step it is important that the mechanical engineer discusses with the RF designer about the cell shape in particular its structural properties and possible issues with heat transfer. It may be the case that the optimal RF design, from a~beam dynamics or RF power view-point, is not feasible to realise mechanically, and the `best' design may be a compromise between optimal RF design and manufacturability. Figure~\ref{fig:fig6} depicts the finalised electromagnetic design of the SwissFEL accelerating structure.

\begin{figure}
    \centering
    \includegraphics[width=1\linewidth]{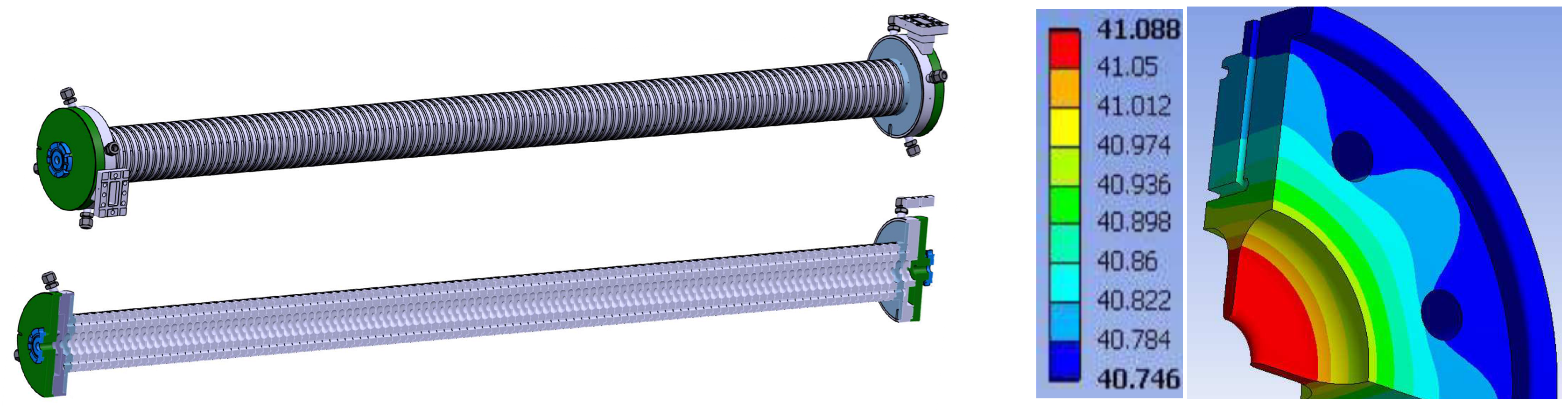}
    \caption{Mechanical design of the full accelerating structure along with the thermal simulations of the single cell~\cite{Raguin2012}.}
    \label{fig:fig7}
\end{figure}

\subsection{Mechanical Design}
\label{Sec:Mech_design}
With the completed electromagnetic design, sometimes referred to as the vacuum volume, the design is then shared in its entirety with the mechanical engineer. It is their job to design a physical structure around this vacuum volume and ensure that the design can be realised mechanically. Therefore, this stage includes important questions like:
\begin{itemize}
    \item What is the difference in operating and machining temperature?
    \item How will the device be joined?
    \item Does the structure require post-fabrication tuning of the cells?
    \item Where will the cooling pipes be located?
\end{itemize}
Firstly, to achieve such machining accuracy, it was important to consider the thermal expansion caused by the difference in temperature of machining and operation of the accelerating structure. Continuing with the SwissFEL accelerating structure example, these structures operate at 40$^\circ$C while their machining is performed at 20$^\circ$C. This equates to several microns for the C-band cells, which exceeds the accuracy requirements. To account for this, the vacuum volume is scaled for the mechanical design so that when machined at 20$^\circ$C, it will expand to the correct dimensions when operating at 40$^\circ$C.

The single cells of the SwissFEL structures were chosen to be brazed together. This choice was due to the high mechanical tolerance requirements as it was decided that the cavities should not require post-assembly tuning. Furthermore, the joint was placed in the centre of the cell to avoid the liquid brazing metal changing the total volume of the cell. The left of Figure~\ref{fig:fig7} depicts the final mechanical design of the accelerating structure.

Finally, it is also important to understand how the mechanical design will be impacted by the~power losses within the cavities that cause local heating. Thermo-mechanical simulations of the cells were performed to determine the cooling pipe requirements. After performing these simulations it was determined that two cooling pipes per quadrant were required to keep the temperature differential across the cell to 0.3$^\circ$C and at the nominal 40$^\circ$C (see right of Fig.~\ref{fig:fig7}).

\begin{figure}
    \centering
    \includegraphics[width=\linewidth]{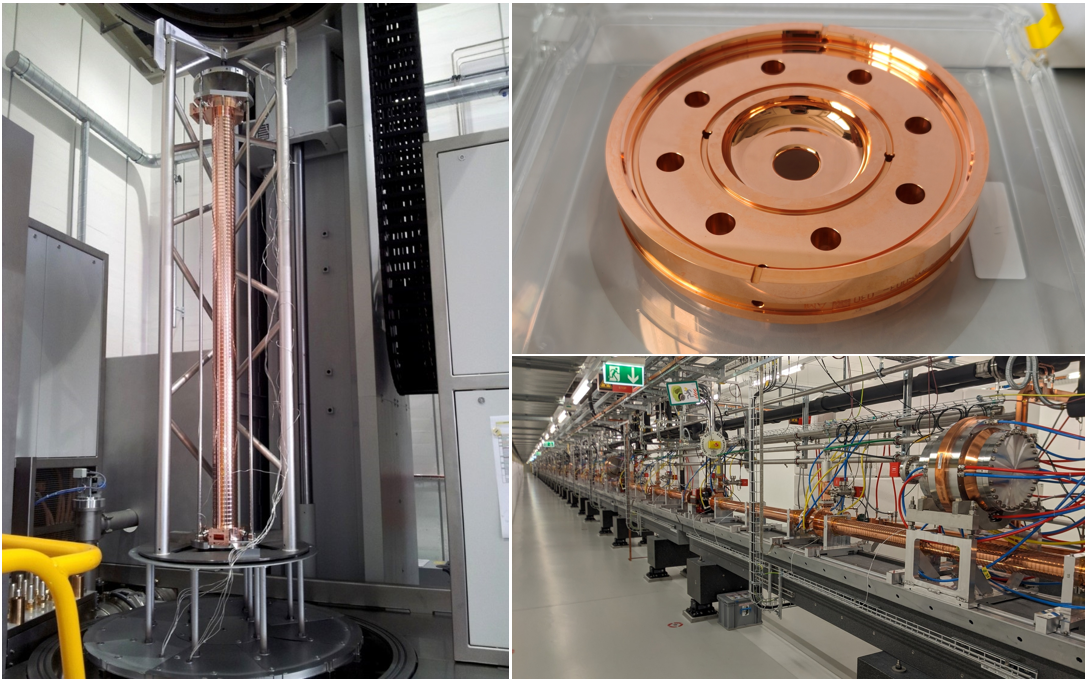}
    \caption{Realised single cell (top right), stacked accelerating structure prepared for brazing (left) and an image of the SwissFEL linear accelerator (right-bottom)~\cite{Raguin2012}.}
    \label{fig:fig4}
\end{figure}

\subsection{Fabrication of the Structure}
For the fabrication of the individual cells and couplers, the raw Cu-OFE pieces started with a pre-turning process machined to a precision of approximately 10~\textmu m. The pieces were then finished to a precision of $\pm$2 \textmu m with an average surface roughness less than 20~nm. For non-cylindrically symmetric components, such as RF couplers, the components were machined with a 5-axis milling machine. The very high machining accuracy was used to remove the need for post-fabrication tuning.

Prior to brazing, each component was baked at 400$^\circ$C to remove surface contaminants. Following this, each of the individual components were then assembled using a KUKA KR 30 HA robotic arm and brazed together. Tight fits between pieces were achieved through inducing a large temperature difference between adjacent pieces. The original precision of the components and the use of brazing was key to producing the structure without the need for post-fabrication tuning of the individual cells. The final tuning of the accelerating structure, as a whole system, was performed using the cooling water temperature.

\subsection{Performance Validation}
The first validation of the accelerating structures came through low power testing of the device. This included two steps:
\begin{itemize}
    \item Scattering (S-) parameter measurements;
    \item Bead-pull Measurement.
\end{itemize}
Such measurements were used to illustrate that the electromagnetic properties of the accelerating structure matched those predicted by the simulations, performed in the initial step. However, the ultimate validation of the structure performance came with the high-power operation and measurements with beam~\cite{Lucas2025}.

\section{Conclusion}

RF systems are critical to the operation of particle accelerators, enabling the acceleration, manipulation and measurement of charged particle bunches. One of the key components is the RF cavity. While the~electromagnetic properties underpin the means of operation of RF cavities, their practical implementation depends on careful mechanical engineering. Therefore, their realisation requires strong collaboration between the mechanical engineer and RF designer.

\section{Acknowledgements}
I would like to acknowledge Marco Pedrozzi whose lectures were the basis of those presented at the~School.

\end{document}